\documentclass[preprint]{aastex}

\slugcomment{Version 3.0 \today}

\shorttitle{Parasitic Light in NGST instruments}
\shortauthors{\dots}

\begin{document}


\title{Parasitic Light in NGST instruments:\\ 
the accuracy of photometric redshifts and the
effect of filter leaks in the visible and near IR camera.}


\author{Stefano Cristiani}
\affil{ ST European Coordinating Facility, European Southern Observatory \\ 
Karl-Schwarzschild-Strasse 2,  D-85748 Garching bei M\"unchen  }

\author{Stephane Arnouts}
\affil{European Southern Observatory\\ Karl-Schwarzschild-Strasse 2,
D-85748 Garching bei M\"unchen } 
%%\email{sarnouts@eso.org}

\author{Robert A.E. Fosbury}
\affil{ ST European Coordinating Facility, European Southern Observatory \\ 
Karl-Schwarzschild-Strasse 2,  D-85748 Garching bei M\"unchen  }

\begin{abstract}
A detailed analysis of NGST multiband photometry applied to the
reference case of the study of high-redshift galaxies has been carried
out with simulations based on galaxy SEDs derived from the currently available
empirical and model templates and on plausible standard filter-sets.

In order to correctly identify star forming galaxies in the redshift
range $5 < z < 9$ and early-type galaxies above $z>2$ and avoid
confusion with other SEDs it is mandatory to have photometric
information in optical bands, besides a standard IR filter-set like 
$~F110 ~F160 ~K ~L ~M$. In particular by adding the
$V606, ~I814$ and $~z-{\rm Gunn}$ filters a good discrimination is obtained
above $z>5$ for star forming galaxies and  
$z>1$ for early-types.
The case for an extension of the NGST wavelength domain to the optical 
range is therefore strongly supported by this analysis.

The effects of leaks in the filter blocking have also been
investigated. In spite of rather pessimistic assumptions (a constant
leak at a level of $10^{-4}$ of the peak transmission over the whole
spectral range or a leak of Gaussian shape placed at 1.5 times the effective
wavelength of the filter with an amplitude of $10^{-3}$ of the peak
transmission and a width of 5\% of the central wavelength)
the effects are not dramatic:
the accuracy of the determination of photometric redshifts 
with a standard $ V606 ~I814 ~z-{\rm Gunn} ~F110 ~F160 ~K ~L ~M $ filter set 
does not significantly deteriorate for a sample limited to $M_{AB} \sim
30$. In the range $5.5 < z <12.5$ a rather high accuracy with a typical
error $|z_{sim}-z_{phot}| \le 0.2$ is achieved.

With the filter blocking achieved by the present standard technology,
extended and continuous spectral coverage appears to be the driving
factor to maximize the scientific output.
\end{abstract}


%%%\keywords{\dots}


\section{Introduction}

\begin{deluxetable}{lcc}
%\tabletypesize{\scriptsize}
\tablecaption{Visible-NIR Camera \label{tbl:NIRcamera}}
\tablewidth{0pt}
\tablehead{
\colhead{ } & \colhead{Optimal}   & \colhead{Bare Bones} 
}
\startdata
$\lambda$ range & $0.6 - 5 \mu m$  &  $ 1.0 - 5.0 \mu m$ \\
Sampling        & $0.03''$         &  $0.03''$  \\
FOV 		& $4' \times 4'$   &  $2' \times 2'$ \\
$\lambda / \Delta\lambda$ & $\leq 10$ &  $\leq 10$ 
\enddata
\end{deluxetable}

The NGST Design Reference Mission (DRM) is a representative science
program elaborated in sufficient detail to aid in the development of functional
requirements for the NGST mission.
One of the main results of the DRM has been the identification of a
large fraction of subprograms (8 out of 12) requiring observations at 
optical wavelengths.
Such an extension poses significant technical challenges.
In particular it requires an efficient blocking of photons at undesired 
wavelengths over a range covering more than three octaves.
In the following we examine the consequences of
leaks in the transmission of filters for a typical programme of
determination of photometric redshifts of high-redshift galaxies. 

\section{Baseline of the Instrument}
Table~\ref{tbl:NIRcamera} shows the main characteristics of the
Visible-NIR camera as conceived by the  Visible and NIR camera
Subcommittee of the ``Ad Hoc Science Working Group'' (ASWG).
Two options are envisaged: an {\it optimal} and a {\it bare bones}
version. More details can be found at the URL \\
{\tt http://www.stecf.org/ngst/instruments.html}.

\begin{deluxetable}{lrcc}
\tablecaption{Properties of the assumed filter set \label{tbl:filters}}
\tablewidth{0pt}
\tablehead{
\colhead{Name}   & \colhead{$<\lambda>$} &
 \colhead{FWHM} & \colhead{${AB}-$Vega} \\
\colhead{ } & \colhead{(\AA) } & \colhead{(\AA) } &
\colhead{correction}
}
%F606   &   6031  &   5831.6 &    2050 & 0.096 \\
%F814   &   8011  &   7862.3 &    1400 & 0.417 \\
%Zgunn  &   9125  &   9066.8 &    1200 & 0.527 \\
%F110   &  11288  &  10493.9 &    5750 & 0.698 \\
%F160   &  16096  &  15798.5 &    3750 & 1.313 \\
%Ks     &  21640  &  21522.1 &    2750 & 1.841 \\
%L      &  37789  &  37497.2 &    5800 & 2.917 \\
%M      &  47657  &  47374.8 &    6400 & 3.380 \\
\startdata
F606   &   6031  &   2050 & 0.096 \\
F814   &   8011  &   1400 & 0.417 \\
z-Gunn &   9125  &   1200 & 0.527 \\
F110   &  11288  &   5239 & 0.698 \\
F160   &  16096  &   3750 & 1.313 \\
Ks     &  21640  &   2750 & 1.841 \\
L      &  37789  &   5800 & 2.917 \\
M      &  47657  &   6400 & 3.380 \\
\enddata
\end{deluxetable}

\section{Band-passes of the filters}
\begin{figure}
\figurenum{1}
\epsscale{0.9}
\plotone{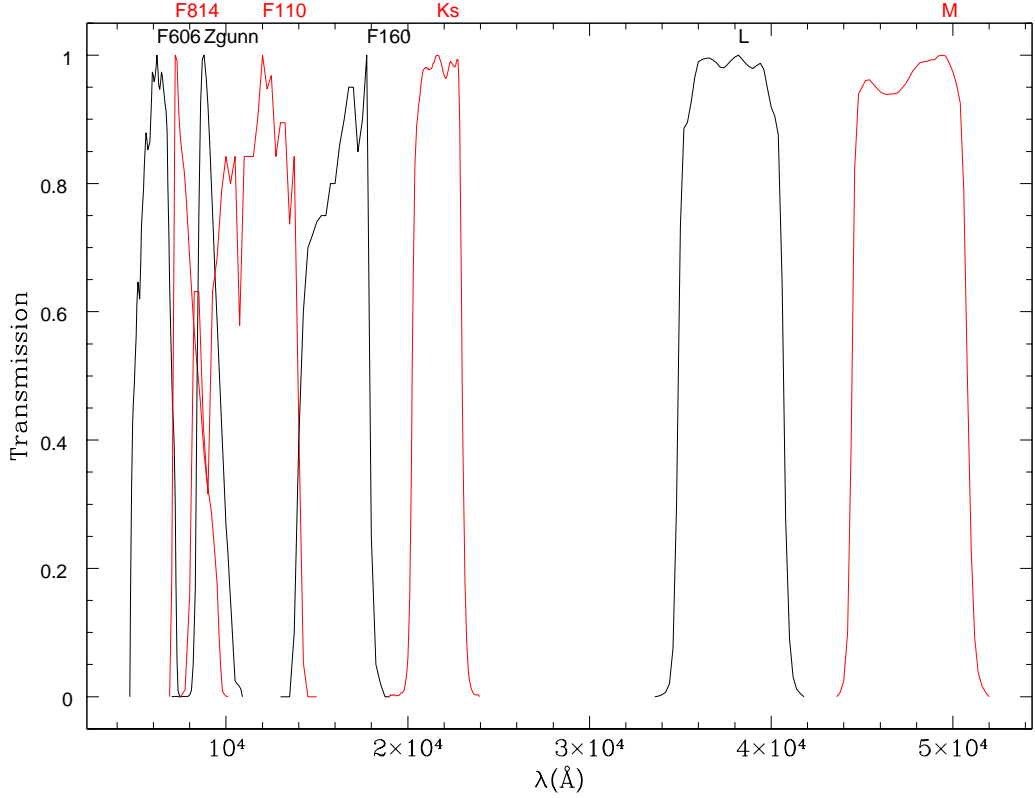}
\caption{The reference filter set used for the simulations.}
\label{fig_filters}
\end{figure}

We have assumed as reference bandpasses for the filters of
the optical-NIR camera the
standard $ V606 ~I814 ~z-{\rm Gunn} ~F110 ~F160 ~K ~L ~M $ set commonly 
used on HST with WFPC2 ($V606 ~I814$) or NICMOS ($F110 ~F160$) and from
the ground (Fig.~\ref{fig_filters}).
Details about the filter properties are given in Table~\ref{tbl:filters}

The actual choice of the filters to be used in space with NGST may be
different but the arguments developed in the following 
do not depend critically on this detail.

\section{Typical SEDs of the sources and the locus of galaxies in
multicolor space} 

\begin{figure}
\figurenum{2}
\epsscale{0.95}
\plotone{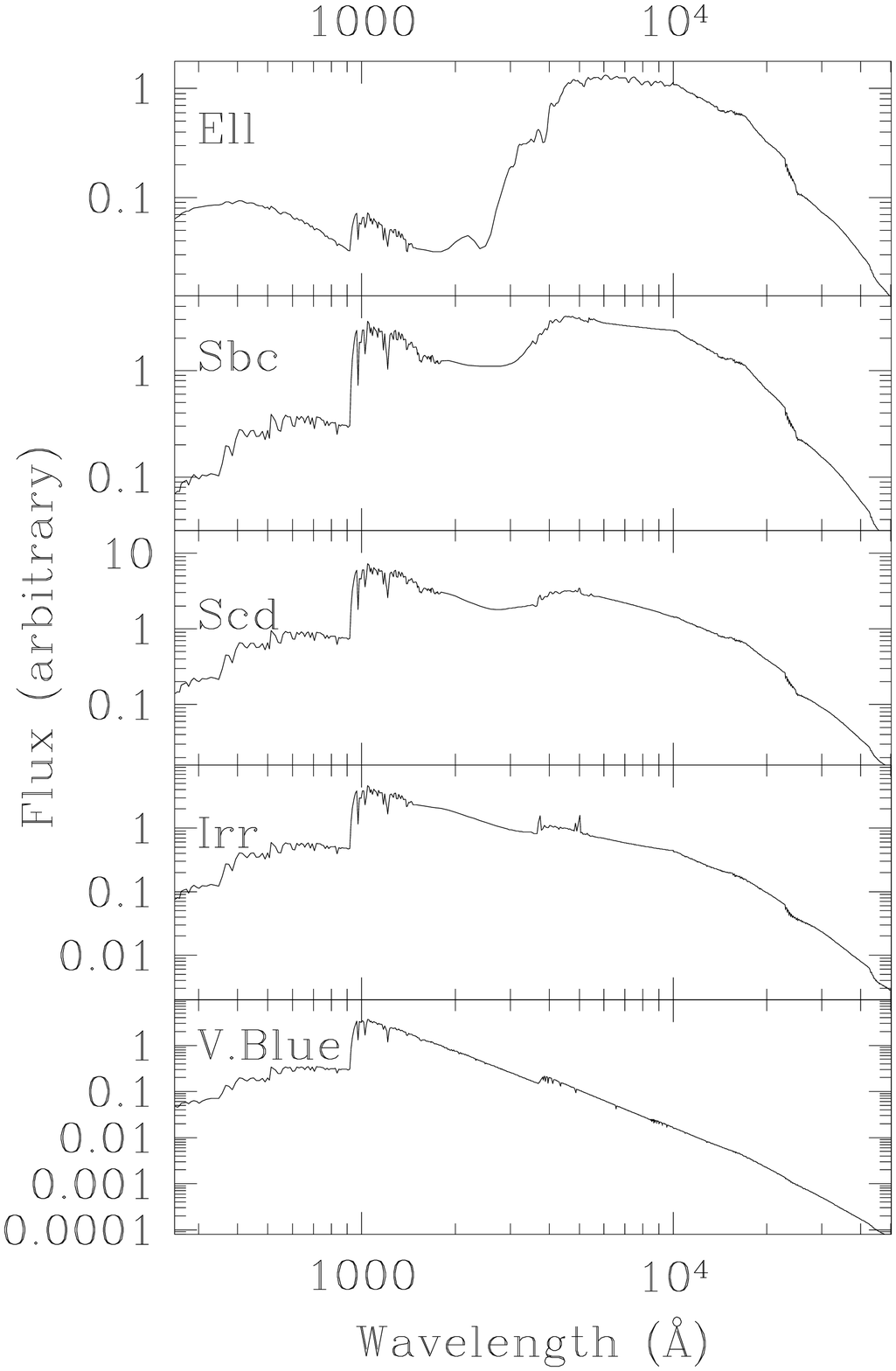}
\caption{The spectral energy distribution of five
template galaxies.}
\label{fig_sed}
\end{figure}

To simulate a range of galaxy SEDs we have adopted the templates of
\citet{coleman80} for a typical elliptical, Sbc, Scd and
irregular galaxy plus a GISSEL \citep{GISSEL} template for a very blue
galaxy (with a constant star formation rate and an age of 0.1 Gyr).
The SEDs are shown in Fig.~\ref{fig_sed} and in the following will be
referred to as extended CWW templates.

\begin{figure}
\figurenum{3}
\epsscale{0.65}
\plotone{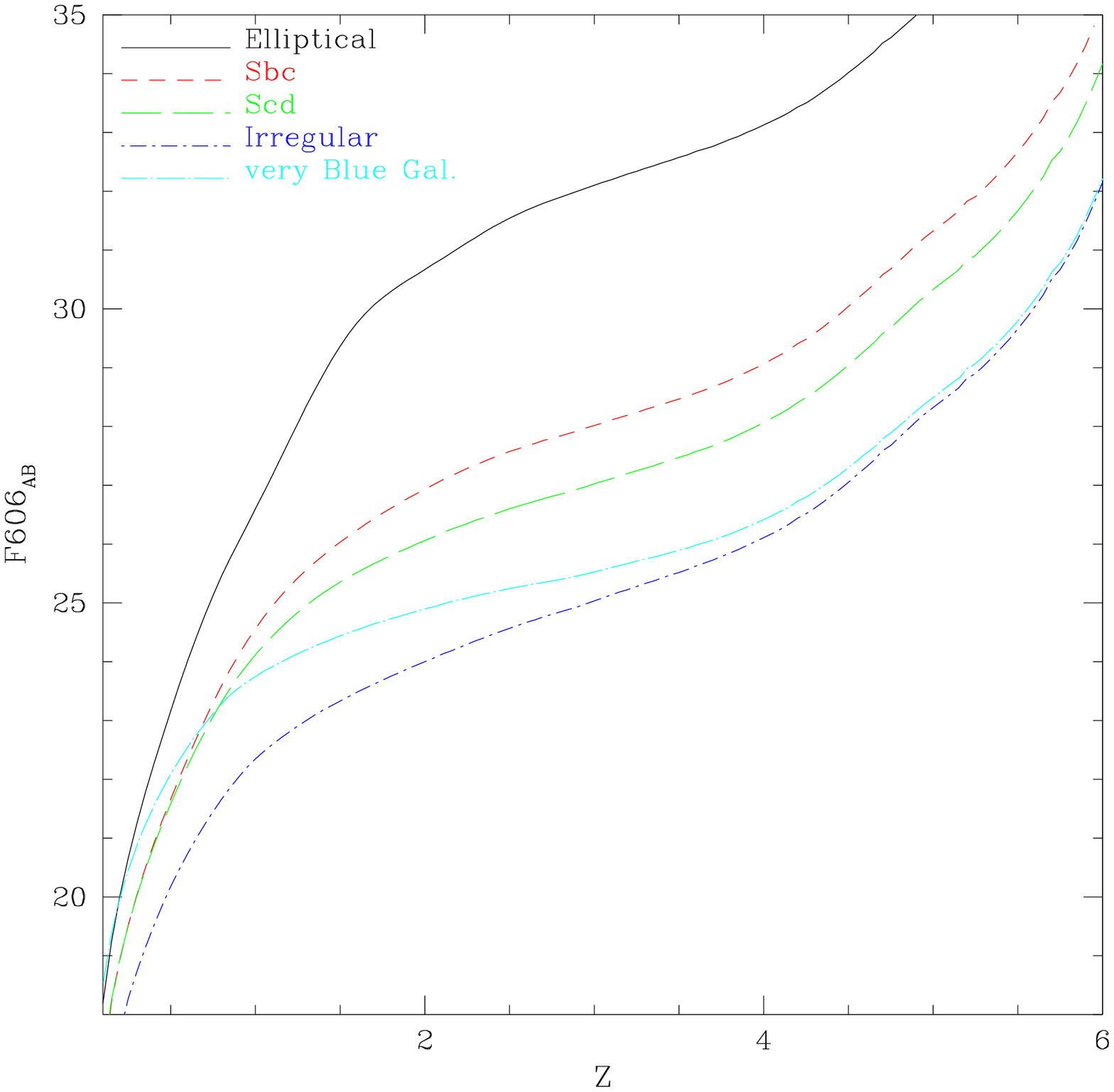} 
\plotone{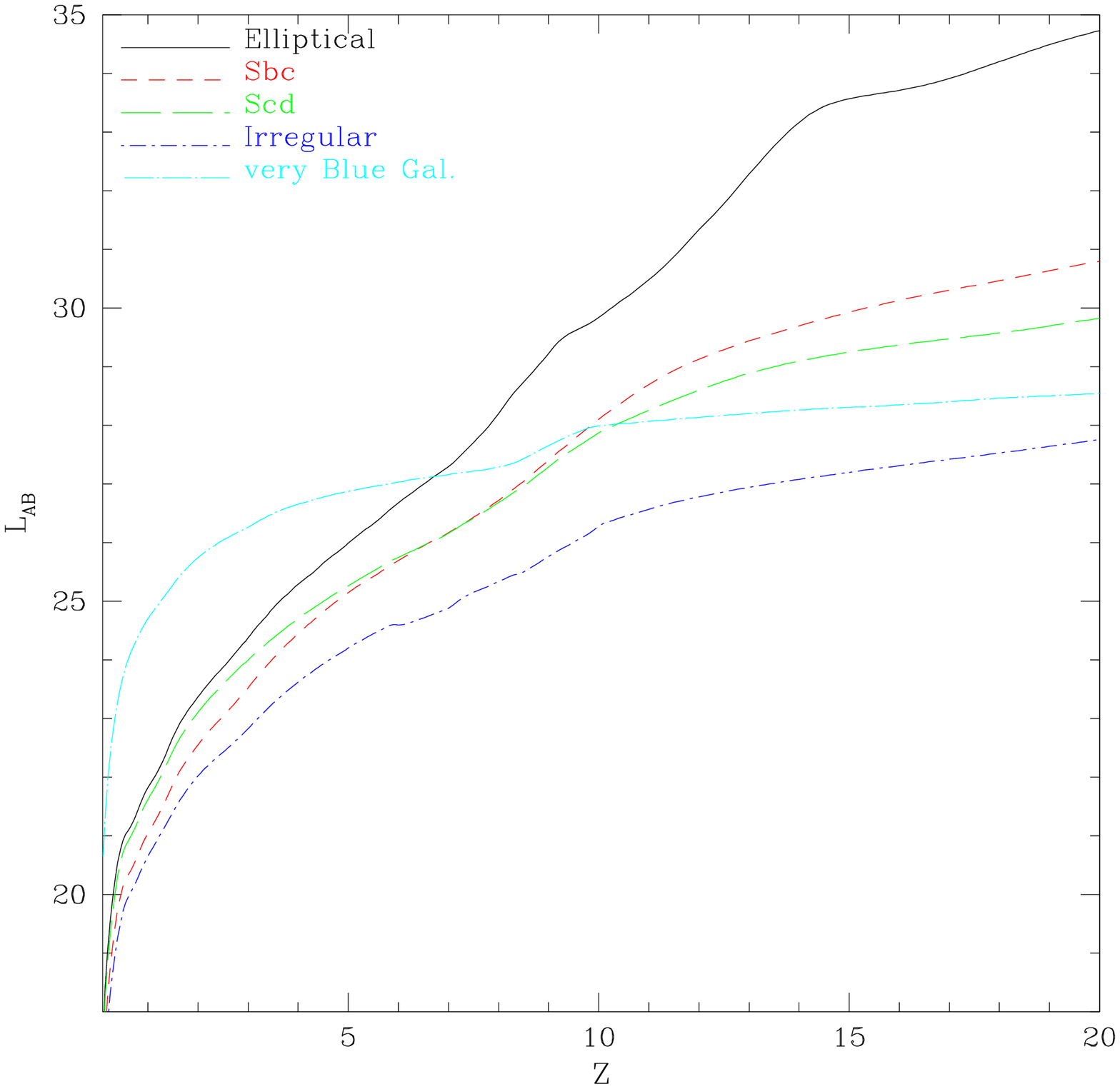}
\caption{Evolution of the apparent magnitude of typical galaxies of
$10^{11} M_{\odot}$.}
\label{fig_appmag}
\end{figure}

Fig.~\ref{fig_appmag} shows the evolution of the apparent magnitude of
typical galaxies of $10^{11} M_{\odot}$ in the $F606$ and $L$ bands
and gives a rough idea of the magnitudes of interest.  It can be
appreciated how an elliptical fades away relatively quickly in the
$F606$ band, while other star-forming galaxies remain - as observed -
significantly brighter.

\begin{figure}
\figurenum{4}
\epsscale{0.65}
\plotone{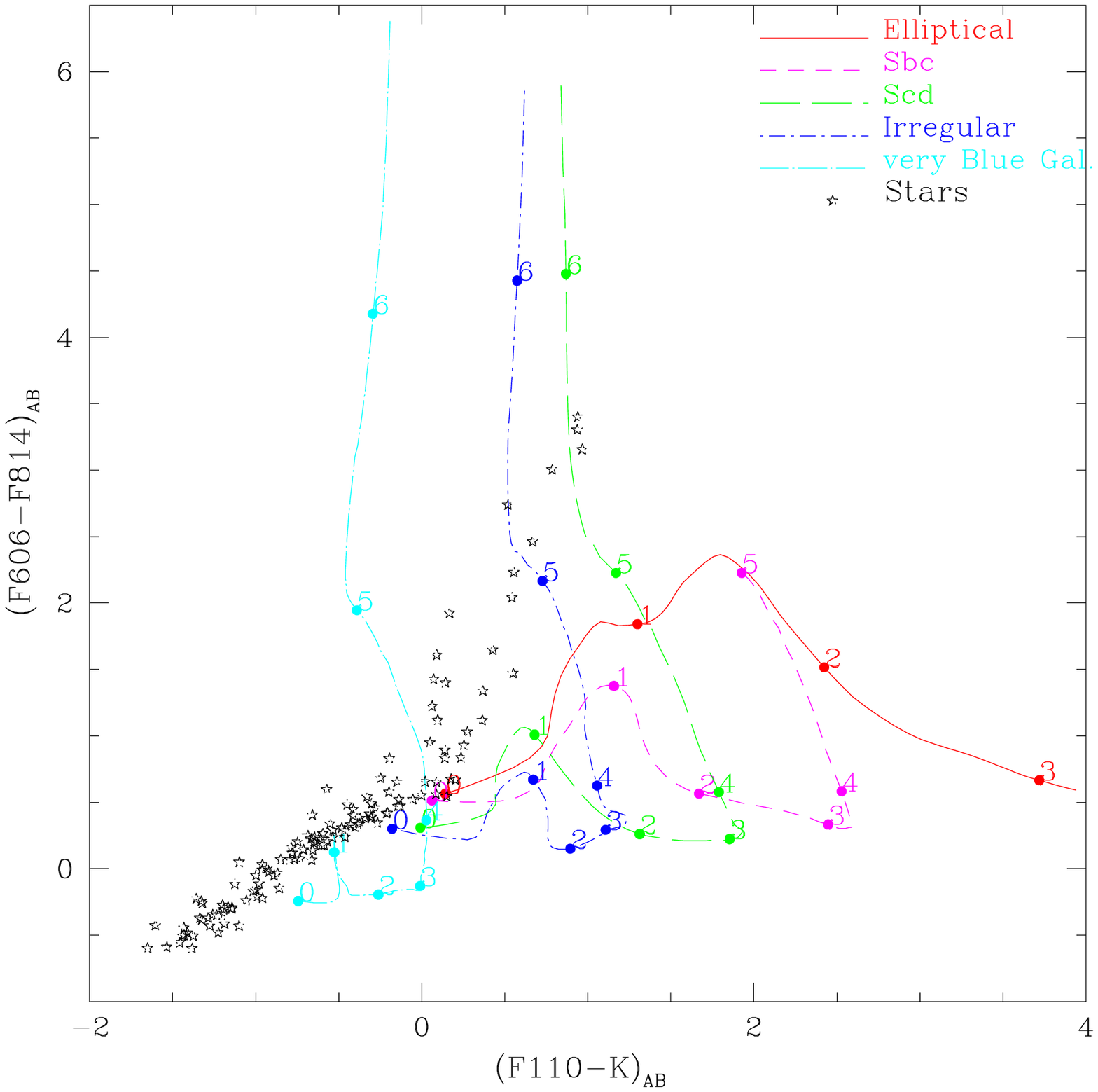}
\plotone{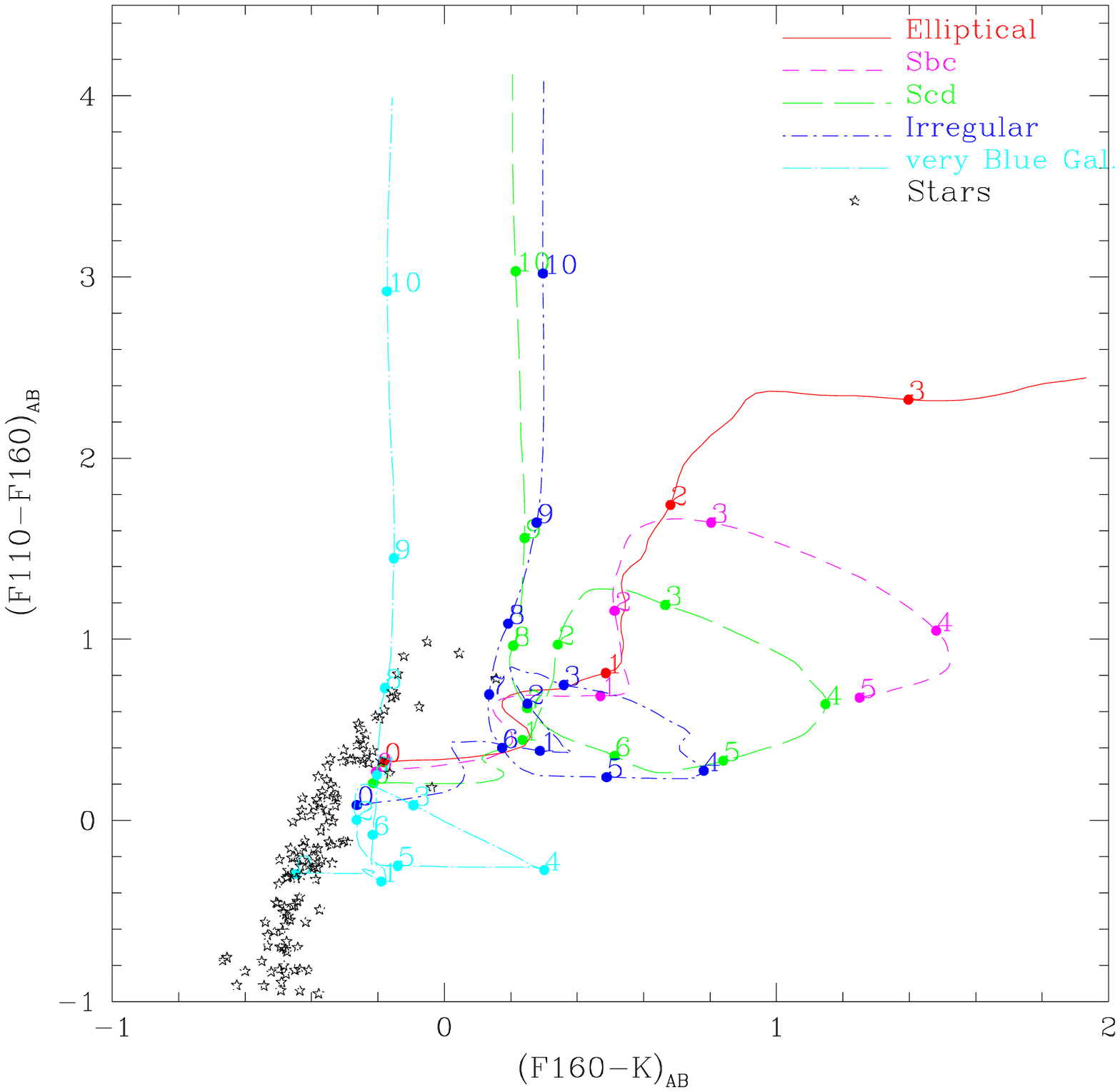}
\caption{Redshift evolution of the colors of galaxies. Upper panel:
$(F606-F818)$ vs. $(F110-K)$. Lower panel:  $(F110-F160)$ vs. $(F160-K)$.}
\label{fig_galloc}
\end{figure}

The upper panel of Fig.~\ref{fig_galloc} shows the evolution of the
colors of the galaxies described in the previous section in a
$(F606-F814)$ vs.  $(F110-K)$ plane. The color tracks are limited to
$z<3.5$ for an elliptical SED and to $z<5$ for Sbc galaxies.  With
this combination of bandpasses it is straightforward to identify
star-forming galaxies with $z>5$ and ellipticals at $z>1$.  The lower
panel shows the same in a $(F110-F160)$ vs.  $(F160-K)$ plane and
demonstrates how, without optical bands, it becomes difficult to isolate
star-forming galaxies at $z<9$.

\section{Photometric redshifts and the effects of ``(red) leaks''}
\begin{figure}
\figurenum{5}
\epsscale{1.0}
\plotone{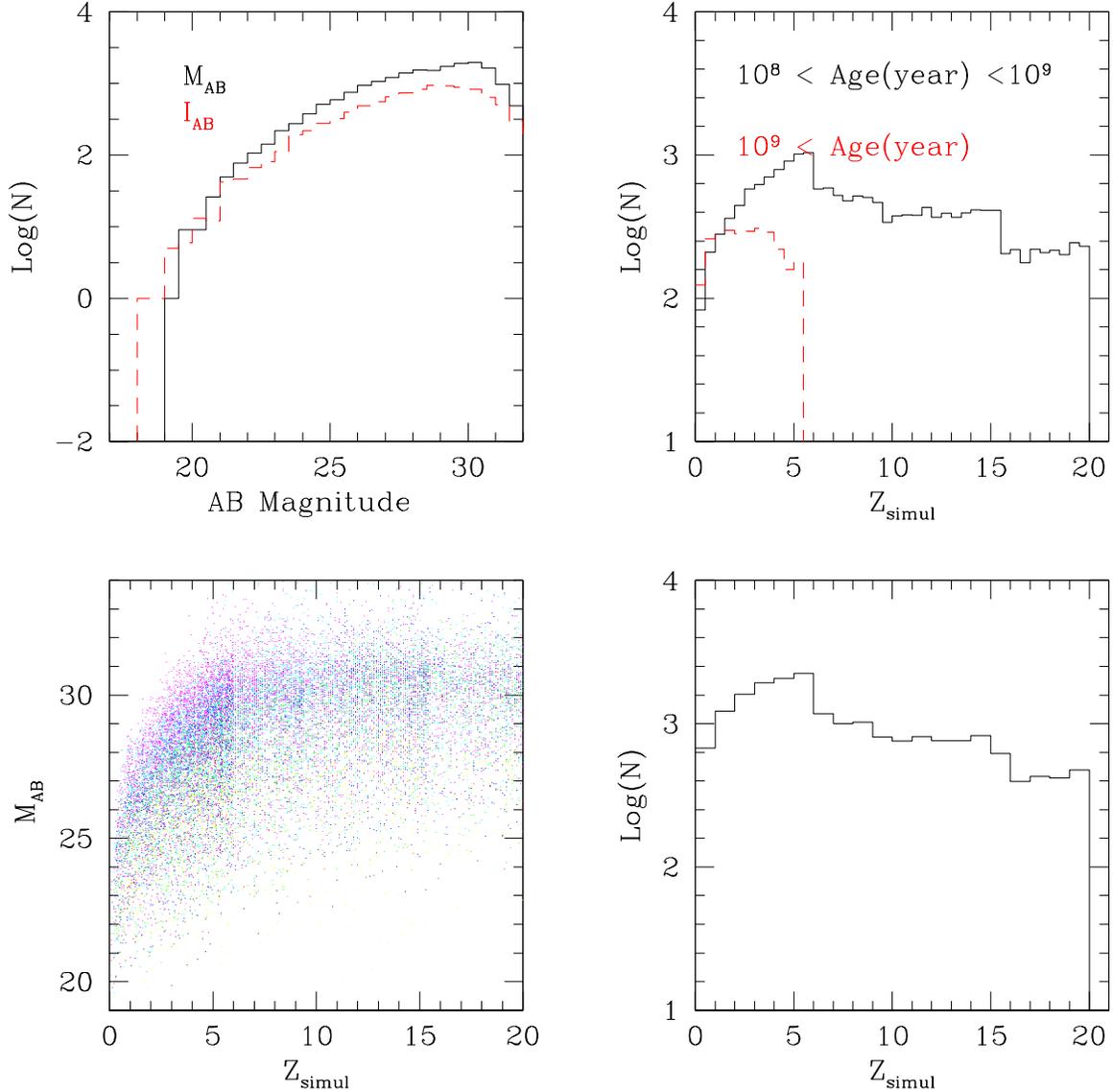}
\caption{Characteristic distributions of the simulation of galaxies
derived from GISSEL models: differential counts as a function of the apparent
magnitude (upper left), redshift distribution (upper and lower right), the
$z-mag$ plane (lower left). In
the upper left panel the continuous and dashed lines refer to the
differential distribution in $M$ and $I$ magnitude, respectively.
In the upper right panel the dashed line refers to galaxies with ages
larger than 1 Gyr, the continuous line to ages between $10^8$ and $10^9$
yr. 
The colors of the dots in the lower left panel (in the electronic version)
correspond to different galaxy types with the same color coding 
of Fig.~\ref{fig_galloc}.
}
\label{fig_simulation}
\end{figure}

In order to quantify the different performance of different sets of
filters and the effects of ``red-leaks'' in a filter configuration
extending from the visible to the infrared we have carried out
simulations of the assignment of photometric redshifts to a sample of
faint galaxies.
To this end, we have tried to generate a simulated photometric catalog
covering a very large range of galaxy spectral properties
and to estimate photometric redshifts from an
economic and independent a set of templates as possible.
Galaxies were simulated in a redshift interval $0<z<20$ according to
GISSEL library.
This spectral synthesis model is controlled by
a number of free parameters. The star formation rate for a galaxy at a 
given age is governed by the assumed $e-folding$ star formation
time-scale $\tau$. Several values of $\tau$ and galaxy ages 
are necessary to reproduce the different observed spectral types.
For the shape of the initial mass function (IMF) we have assumed a
Salpeter law, which turns out not to be a critical choice.
In addition to the GISSEL parameterization, we have added the
internal reddening of each galaxy by applying the observed
attenuation law for local starburst galaxies by Calzetti et al.
(1997), with the $E_{B-V}$ normalization varying between 0 and
0.2 mag.
We have also included the Lyman absorption produced by the
intergalactic medium  as a function of redshift.
More details on this type of simulations can be found in 
\citet{arnouts99}.
Fig.~\ref{fig_simulation} shows the galaxy counts as a function of the
apparent magnitude, the age and redshift distribution, the $z-mag$
plane characterizing the simulation.

Magnitudes and colors have been derived
from the simulated SEDs by
convolution with the filter passbands described in Sect.3.
Measurement errors have been added, according to the NGST mission simulator
({\tt http://www.ngst.stsci.edu/nms/main/}), \\
 assuming an exposure time
of $10^4$s.

From this set of ``simulated observed colors''
photometric redshifts have been derived using
a standard procedure of
$\chi^2$ optimization, comparing the observed fluxes with errors
to the extended CWW set of templates described
in Sect.~4.

The $\chi^2$ has been computed as:
\begin{equation}
\chi ^2 = \sum_i \left[ {F_{observed,i}- A \cdot F_{template,i} \over
\sigma_i} \right]^2
\end{equation}
where $F_{observed,i}$ is the flux observed in a given filter $i$,
$\sigma _i$ is its uncertainty, $F_{template,i}$ is the flux of the
template in the same filter and the sum is over the various filters.
 The template fluxes have been normalized to the observed ones with
the factor $A$ minimizing the $\chi^2$ value : 
\begin{equation}
A = \sum_i \left[ {F_{observed,i} \cdot  F_{template,i} \over
\sigma_i^2} \right]  /  \sum_i \left[ {F_{template,i}^2 \over
\sigma_i^2} \right]
\end{equation}

\begin{figure}
\figurenum{6}
\epsscale{1.0}
\plotone{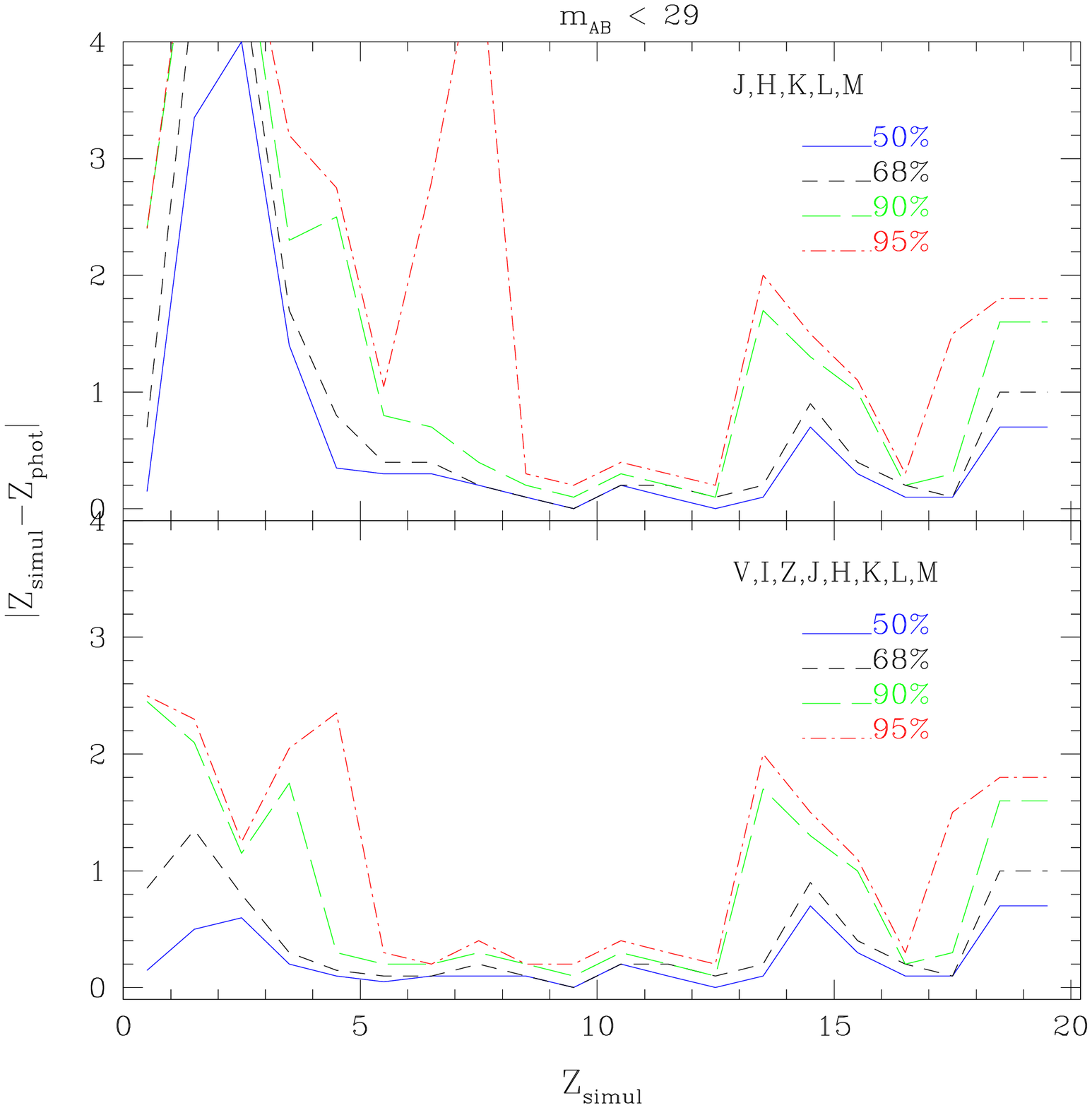}
\caption{Comparison between the accuracy of the photometric redshift
estimates with a filter set including IR and optical bands (upper panel
 $ V606 ~I814$ $~z-{\rm Gunn} ~F110(J)  ~F160(H) ~K ~L
~M $) and a filter set limited to infrared bands (lower panel).
The dot-dashed, long-dashed, short-dashed and continuous
lines correspond to the
95, 90, 68 and 50 percentile of the absolute deviation
$|z_{sim}-z_{phot}|$, respectively. 
The galaxy sample is limited to $M_{AB}<29$.}
\label{fig_optIR29}
\end{figure}
Fig.~\ref{fig_optIR29} ~shows the comparison between photometric redshift
estimates obtained with a $ V606 ~I814$ $~z-{\rm Gunn} ~F110 ~F160 ~K ~L
~M $ filter set and a filter set limited to the infrared bands.
As suggested in the previous section
in order to correctly identify star forming galaxies in the redshift
range $5 < z < 9$ and avoid
confusion with other SEDs it appears mandatory to have, besides the IR
filters, photometric information in the optical bands.
In the following we will adopt the set 
$ V606 ~I814 ~z-{\rm Gunn} ~F110 ~F160 ~K ~L ~M $ as standard.

\begin{figure}
\figurenum{7}
\epsscale{1.0}
\plotone{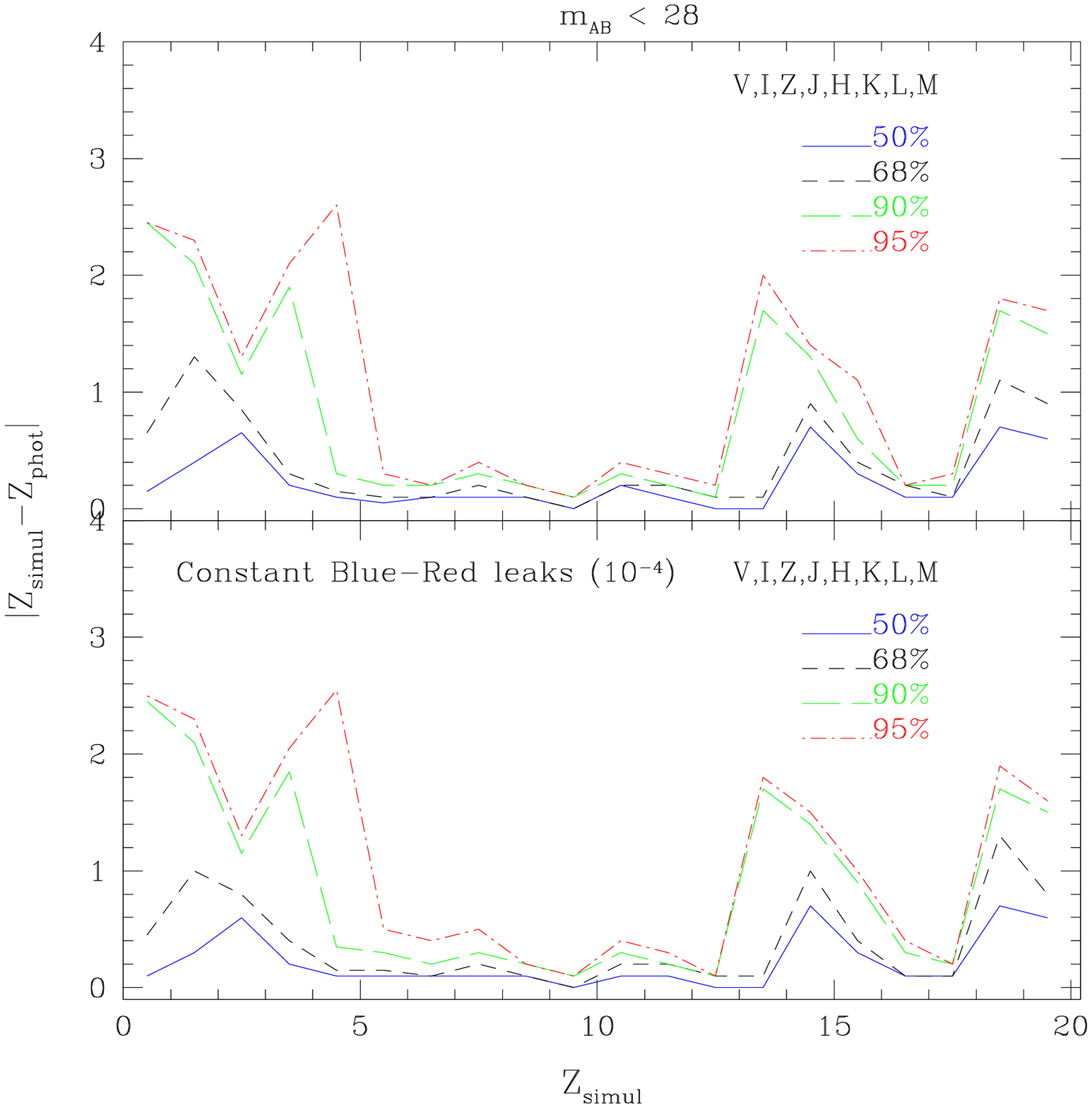}
\caption{Typical errors of the photometric redshifts with a
$V606 ~I814 ~z-{\rm Gunn} ~F110(J) ~F160(H) ~K ~L ~M$ filter set.
The upper panel shows the result obtained assuming only the standard
photometric error, the lower panel includes the effects of a constant
leak at a level of $10^{-4}$ of the peak transmission over the whole
spectral range. 
The dot-dashed, long-dashed, short-dashed and continuous
lines correspond to the
95, 90, 68 and 50 percentile of the absolute deviation
$|z_{sim}-z_{phot}|$, respectively. 
The galaxy sample is limited to $M_{AB}<28$.}
\label{fig_zphot_v_cl28}
\end{figure}
\begin{figure}
\figurenum{8}
\epsscale{1.}
\plotone{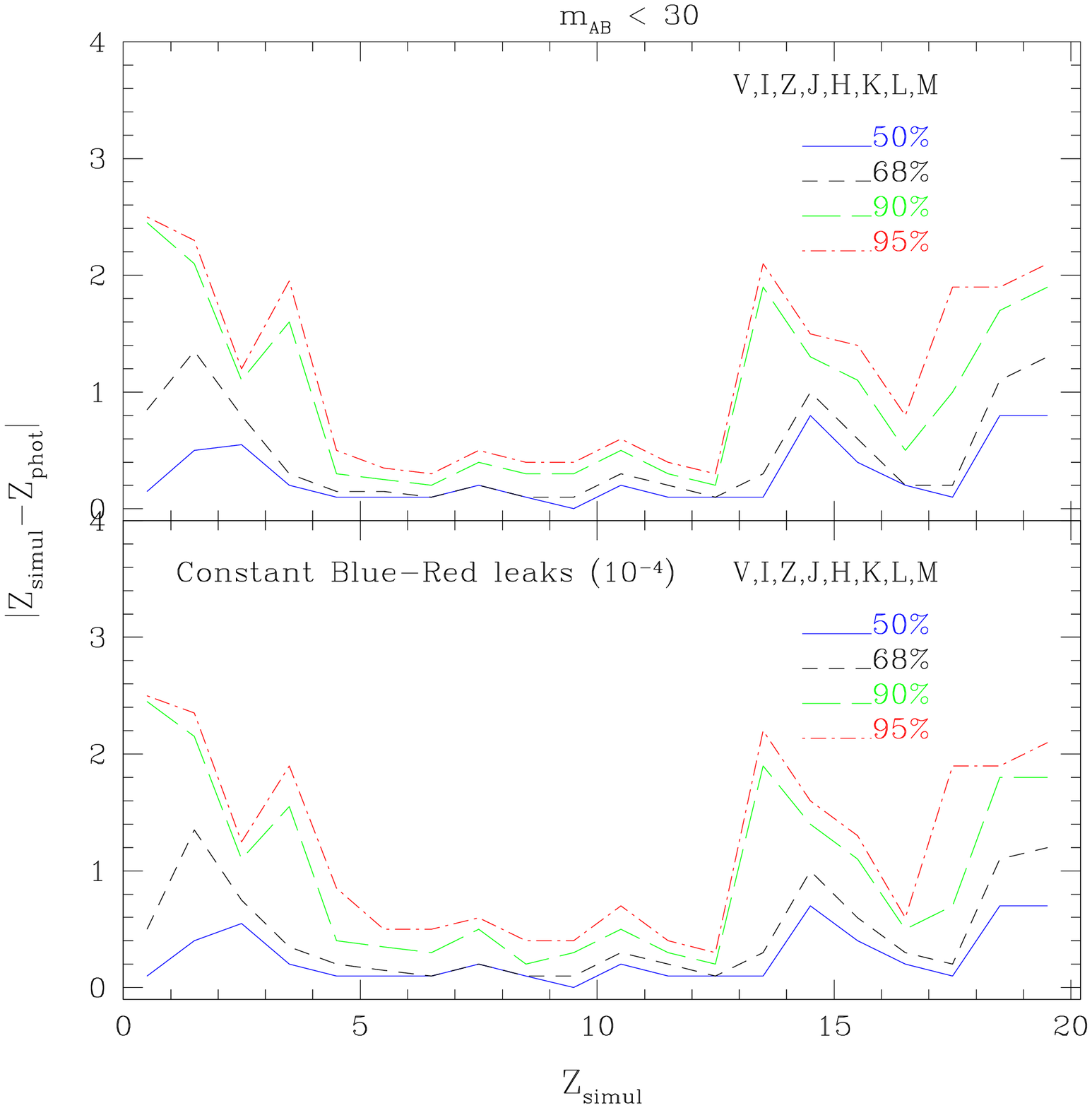}
\caption{
Same as Fig.~\ref{fig_zphot_v_cl28}, but down to $M_{AB}=30$.}
\label{fig_zphot_v_cl30}
\end{figure}
\begin{figure}
\figurenum{9}
\epsscale{1}
\plotone{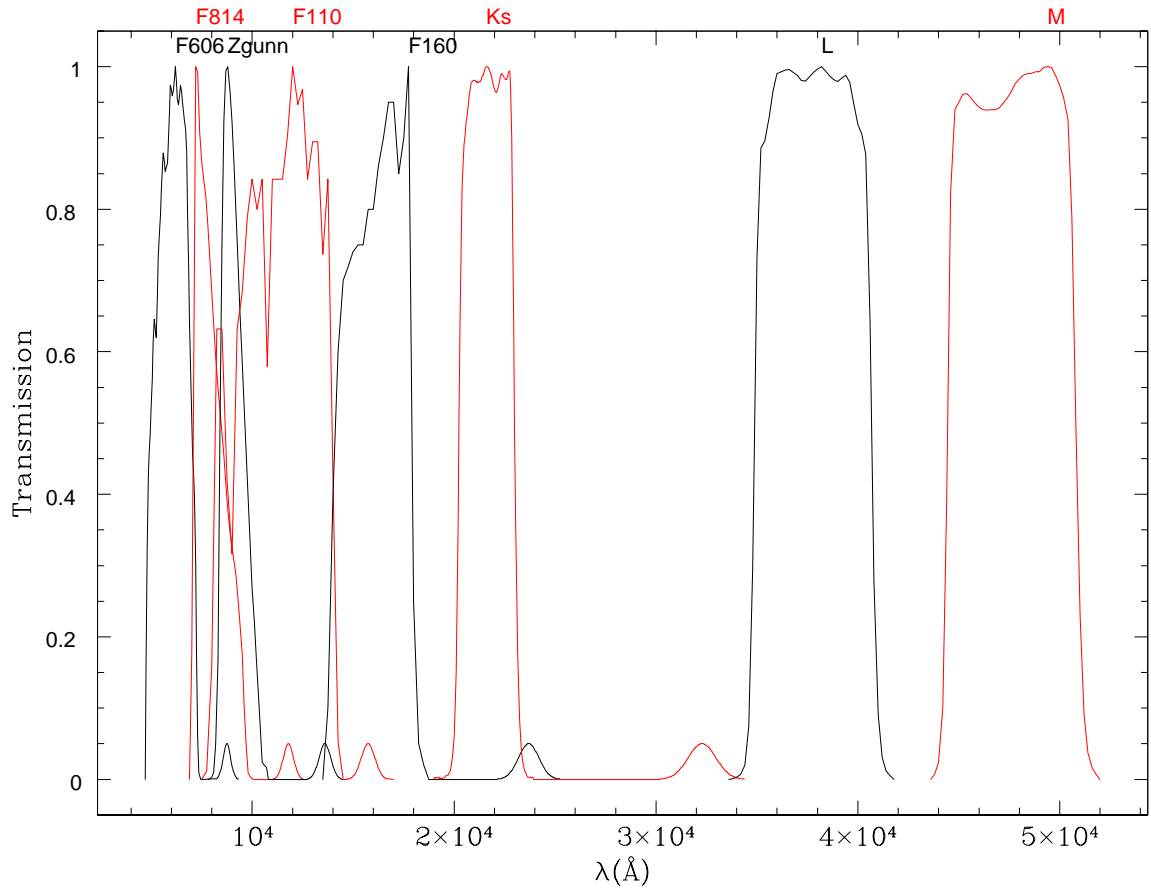}
\caption{
The response of the $V606 ~I814 ~z-{\rm Gunn} ~F110 ~F160 ~K ~L ~M$ filter
with the addition of a leak of Gaussian shape placed at 1.5 times the effective
wavelength of the filter with an amplitude of $10^{-3}$ of the peak
transmission and a width of 5\%
of the central wavelength.}
\label{fig_filter_gl}
\end{figure}
\begin{figure}
\figurenum{10}
\epsscale{1.0}
\plotone{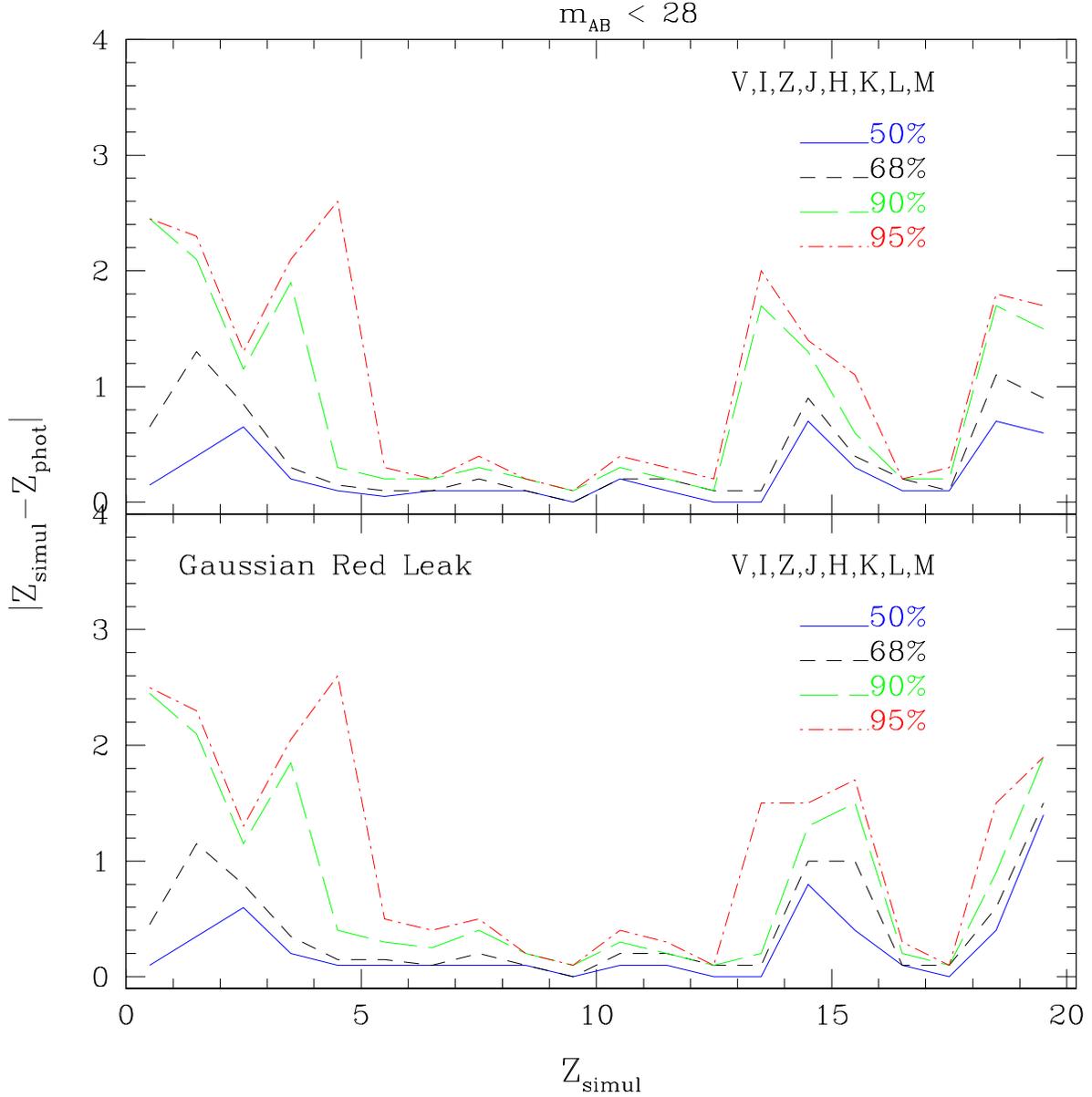}
\caption{Typical errors of the photometric redshifts with a
 $ V606 ~I814 ~z-{\rm Gunn} ~F110 ~F160 ~K ~L ~M $ filter set.
The upper panel shows the result obtained assuming only the standard
photometric error, the lower panel includes the effects of a leak of
Gaussian shape placed at 1.5 times the effective wavelength of the
filter with an amplitude of $10^{-3}$ of the peak transmission and a
width of 5\% of the central wavelength.  
The dot-dashed, long-dashed, short-dashed and continuous
lines correspond to the 95, 90, 68 and 50 percentile of the
absolute deviation $|z_{sim}-z_{phot}|$, respectively.
The galaxy sample is limited to $M_{AB}<28$.}
\label{fig_zphot_v_gl28}
\end{figure}
\begin{figure}
\figurenum{11}
\epsscale{1.}
\plotone{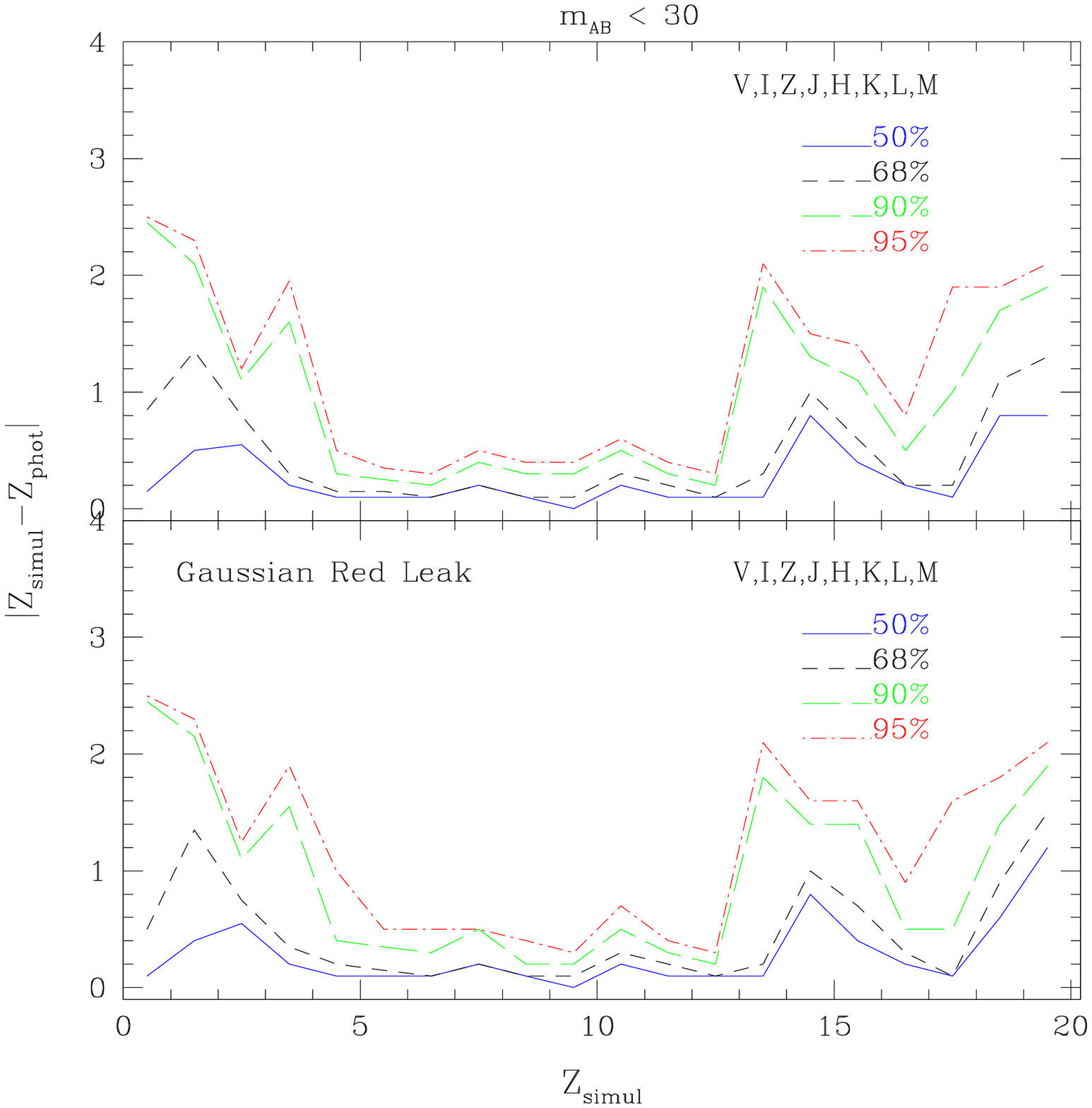}
\caption{
Same as Fig.~\ref{fig_zphot_v_gl28} but down to $M_{AB}=30$.}
\label{fig_zphot_v_gl30}
\end{figure}

In order to explore the effects of filter leaks the
following approach was adopted:
\begin{enumerate}
\item{two types of leaks have been foreseen: a constant leak at a
level of $10^{-4}$ of the peak transmission over the whole spectral
range and a leak of Gaussian shape placed at 1.5 times the effective
wavelength of the filter with an amplitude of $10^{-3}$ of the peak
transmission and a width of 5\%
of the central wavelength. The effect of Gaussian leaks on the
filter transmission is shown in Fig.~\ref{fig_filter_gl}
}
\item{The leaks have been included in the system spectral response and
new magnitudes and colors have been produced with the GISSEL models}.
\item{The new colors have been used to produce photometric redshifts 
with the extended CWW templates. No knowledge of the leaks was used
in this step, i.e. the system responses assumed in the z-phot estimation
are those of Fig.~\ref{fig_filters}, without leaks.}
\end{enumerate}

The typical errors deriving from photometric uncertainties and
unaccounted leaks are illustrated in
Fig.~\ref{fig_zphot_v_cl28}-\ref{fig_zphot_v_cl30} and
\ref{fig_zphot_v_gl28}-\ref{fig_zphot_v_gl30}.
The dot-dashed, long-dashed, short-dashed and continuous
lines correspond to the
95, 90, 68 and 50 percentile of the absolute deviation
$|z_{sim}-z_{phot}|$, respectively.
It can be seen that errors are  in all cases relatively large at 
$z \le 5$. This is an obvious effect due to the lack of bandpasses
bluer than $V_{606}$, and the consequent difficulty to identify
the Lyman break at relatively low redshift and distinguish it, for
example, from the 4000 \AA\ break at lower redshift.

The effect of the leaks, despite the rather pessimistic
assumptions, is in general a moderate
increase of the error level.
For example for $z=11$ and $M_{AB} < 30$ the 68 percentile corresponds 
to $|z_{sim}-z_{phot}| = 0.23 $ with the standard photometric errors, while 
it increases at  $|z_{sim}-z_{phot}| = 0.25$ with the unaccounted
constant leak at a level of $10^{-4}$ of the peak transmission over the whole
spectral range. No qualitative change of the accuracy is observed when 
the leaks are introduced.
No relevant difference is also found between the continuous and the
Gaussian leak cases.
Again for $z=11$ and $M_{AB} < 30$ the 68 percentile corresponds 
to $|z_{sim}-z_{phot}| = 0.27$ when the leak of Gaussian shape is added.
In the range $5.5 < z <12.5$ a rather good accuracy is achieved with
a typical error $|z_{sim}-z_{phot}| \le 0.2$ in all cases down to
$M_{AB} \sim 30$.
The situation deteriorates slightly around $z \simeq 15$, but the
result is due mainly to the insufficient sampling of the SEDs between
the $K_s$ and the $M$ filters, in particular to the ``hole'' at 2.9
$\mu$m.

The importance of the leaks, relative to the uncertainty due to
photometric errors, decreases with decreasing flux. At $M_{AB} < 30$
and $z=11$ $|z_{sim}-z_{phot}| = 0.18, 0.23, 0.23$ for the no-leak,
constant-leak and Gaussian leak case, respectively.

\section{Conclusions}

A detailed analysis of NGST multiband photometry applied to the
reference case of the study of high-redshift galaxies has been carried
out with simulations based on galaxy SEDs derived from the currently available
empirical and model templates and on plausible standard filter-sets.

In order to correctly identify star forming galaxies in the redshift
range $5 < z < 9$ and early-type galaxies above $z>2$ and avoid
confusion with other SEDs it is mandatory to have photometric
information in optical bands, besides a standard IR set like 
$~F110 ~F160 ~K ~L ~M$. In particular by adding the
$V606, ~I814$ and $~z-{\rm Gunn}$ filters a good discrimination is obtained
above $z>5$ for star forming galaxies and  
$z>1$ for early-types.
The case for an extension of the NGST wavelength domain to the optical 
range is therefore strongly supported by this analysis.

The effects of leaks in the filter blocking have also been
investigated. In spite of rather pessimistic assumptions (a constant
leak at a level of $10^{-4}$ of the peak transmission over the whole
spectral range or a leak of Gaussian shape placed at 1.5 times the effective
wavelength of the filter with an amplitude of $10^{-3}$ of the peak
transmission and a width of 5\% of the central wavelength)
the effects are not dramatic:
the accuracy of the determination of photometric redshifts 
with a standard $ V606 ~I814 ~z-{\rm Gunn} ~F110 ~F160 ~K ~L ~M $ filter set 
is not significantly deteriorated for a sample limited to $M_{AB} =
28$. In the range $5.5 < z <12.5$ a rather good accuracy a typical
error $|z_{sim}-z_{phot}| \le 0.2$ is achieved.

As standard specifications for leaks are $ < 10^{-4}$ and as 
any larger leaks could be measured once the filters are produced, it 
does not look as though leaks constitute a serious problem,
at least for the particular application described here.
The interpretation of fields containing highly reddened sources may,
however, be more seriously compromised.
Extended and continuous spectral coverage appears to be the driving
factor to maximize the scientific output of the NGST camera.

\section{Acknowledgments}
\acknowledgments

We are grateful to A.Moorwood for enlightening discussions about the
performances of IR filters and detectors and a critical reading of the 
manuscript.

\clearpage

\end{document}